\documentclass[aps,prb,groupedaddress,showpacs]{revtex4}

\usepackage{graphicx}

\begin{document}


\title{Localization in a strongly disordered system: A perturbation approach}


\author{Marco Frasca}
\email[e-mail:]{marcofrasca@mclink.it}
\affiliation{Via Erasmo Gattamelata, 3 \\
             00176 Roma (Italy)}


\date{\today}

\begin{abstract}
We prove that a strongly disordered two-dimensional system localizes with a 
localization length given analytically. We get a scaling law with a critical exponent is $\nu=1$
in agreement with the Chayes criterion $\nu\ge 1$.
The case we are considering is for off-diagonal disorder.
The method we use is a perturbation approach holding in the limit of an infinitely large
perturbation as recently devised and the Anderson model is considered with a Gaussian
distribution of disorder. The localization length diverges when energy goes to zero 
with a scaling law in agreement to numerical and theoretical expectations. 
\end{abstract}

\pacs{72.15.Rn, 72.20.Ht, 71.30.+h}

\maketitle


Since the discovery of the existence of a metal-insulator transition by Anderson \cite{and}
originating by a significant presence of disorder in a sample of material, a large part
of activity in condensed matter physics pertained the deep understanding of this effect
originating from the localization of the conduction electrons \cite{mck}.

Several breakthroughs as the scaling theory \cite{lic} appeared that improved the understanding
of the metal-insulator transition. Besides, weak localization theory made clear the effect of
higher order corrections to the scaling theory itself \cite{alt}. 
More recently, the interplay between disorder and
electron-electron interaction has been clarified \cite{fin}. 

As in all phase transitions, also this case depends critically on the dimensionality of the
system. Indeed, it is well-known that for $d\le 2$ the system always localizes and so it is
always in an insulating phase. This result has been challenged by recent experimental findings
\cite{kra1,kra2} showing a 2D metal-insulator transition. This effect cannot be explained
by the localization theory of Anderson and the scaling theory. A satisfactory explanation
has been recently found numerically \cite{bac} through the effect of electron-electron
interaction becoming increasingly important as the density of the two-dimensional electron
gas is lowered producing a rich phase diagram in this limit.

Notwithstanding the increasing relevance that electron-electron interaction has assumed
in recent years, there are a lot of situations for disordered crystals where the Anderson model
proves to be a satisfactory description of the physics while leaving opened several problems.
Current studies point toward a proper understanding of the scaling of the localization
length with some specific parameters of the crystal as could be the energy. This scaling
is characterized by a critical exponent $\nu$ that has been obtained both by experiments
\cite{ex1,ex2,ex3} and numerically \cite{mck,nu1,nu2,nu3} giving results 
for $\nu$ ranging from 1 to 1.6 for 3D systems. 

As it is clear from the above discussion, the lack of an analytical approach prevented
an analysis of the Anderson model in the regime of strong disorder. This is a rather
general situation in physics. Our aim in this paper is to apply a recent devised perturbation
method conceived to treat this kind of problems. This method has been initially applied to
quantum mechanics \cite{fra1,fra2,fra3} and recently to quantum field theory \cite{fra4}.
Applications in mesoscopic physics were also devised \cite{fra5,fra6,fra7}.

The method relies on the duality principle in perturbation theory \cite{fra3} that, using
the arbitrariness in the choice of a perturbation, yields a dual series to the weak perturbation
one. The dual series should be intended as the one holding for the
development parameter going to infinity. As we showed recently, for a time-independent
quantum mechanical problem, the dual perturbation series is given by the Wigner-Kirkwood
expansion \cite{fra8}. We will use this latter result to obtain the localization length
for a strongly disordered two-dimensional system in a closed form.

The result we obtain is that the localization length scales as $1/E$ being $E$ the energy
and then we have $\nu=1$ at the leading order
for off-diagonal disorder. 
Both the localization length and the propagator are obtained
analytically assuming Gaussian the distribution of the disorder. It is interesting to note
the relevance that the Wigner-Kirkwood expansion can attain beyond the already known applications
in statistical mechanics. We assume this result to hold in two dimension.
Recent numerical and theoretical results \cite{ro1,ro2,rai} point out a scaling law
with a divergence at $E=0$ but with $\nu$ ranging from 0.3 to 0.6. We note anyhow that
our leading order is an Anderson model with pure disorder, due to the fact
that we are taking a very strongly disordered system, 
and this difference should be
expected. To exploit this point we compute the first order correction. In any case
it should be emphasized that our result agrees with the Chayes criterion 
as it should be $\nu\ge 1$ \cite{cha}.

In order to make this paper self-contained we work out the duality principle in perturbation
theory directly on the model of a single electron in a disordered material. We consider
the Hamiltonian for a two dimensional electron gas
\begin{equation}
    H=\int d^2x \left[\frac{\hbar^2}{2m}\nabla\Psi^\dagger({\bf x})\nabla\Psi({\bf x})+\Psi^\dagger({\bf x})V({\bf x})\Psi({\bf x})\right]
\end{equation}
being $\Psi({\bf x})$ the field of the electron, while we take for the random potential $V({\bf x})$ a Gaussian distributed potential having
\begin{equation}
    \langle V({\bf x}_i)\rangle = 0
\end{equation}
and
\begin{equation}
    \langle V({\bf x}_i)V({\bf x}_j)\rangle = u_0^2\delta_{ij}
\end{equation}
having the potential defined on different lattice sites. 
We give explicitly the probability distribution defined as
\begin{equation}
    Z(V)=Ne^{-\frac{V(x)^2}{2u_0^2}}
\end{equation}
that is defined on the sites. $N$ is a normalization constant. This means that a functional integral for computing averages turns into an ordinary multiple integral due to the lattice.
This model can be easily reduced
to the Anderson model with off-diagonal disorder giving the same physics. 

We can give explicitly $u_0$ by
the properties of the electron gas in the material by the relation \cite{mck}
\begin{equation}
    u_0^2=\frac{\hbar}{2\pi\tau n(E)}
\end{equation}
being $\tau$ the mean free time and $n(E)$ the density of states that in our two dimensional
case is $n(E)=mA/\pi\hbar^2$ being $A$ the area and $m$ the electron mass.

Weak perturbation theory is straightforwardly applied to this Hamiltonian \cite{dic}
giving weak localization \cite{alt}. This can be obtained by writing down the
Schr\"odinger equation for the above Hamiltonian as
\begin{equation}
    i\hbar\frac{\partial\Psi}{\partial t}=-\frac{\hbar^2}{2m}\Delta_2\Psi + V({\bf x})\Psi
\end{equation}
and applying the Dyson series assuming the random potential as a weak perturbation
giving straightforwardly for the unitary evolution operator
\begin{eqnarray}
      U(t,t_0) &=& U_0(t,t_0)\left[I-\frac{i}{\hbar}\int_{t_0}^t dt'U_0^{-1}(t',t_0)V({\bf x})U_0(t',t_0)\right. \\ \nonumber
      &-& \left.\frac{1}{\hbar^2}\int_{t_0}^t dt' \int_{t_0}^{t'} dt''
	    U_0^{-1}(t',t_0)V({\bf x})U_0(t',t_0)U_0^{-1}(t'',t_0)V({\bf x})U_0(t'',t_0)+\ldots\right]
\end{eqnarray}
being $U_0$ the solution of the equation
\begin{equation}
    i\hbar\frac{\partial U_0}{\partial t}=-\frac{\hbar^2}{2m}\Delta_2 U_0
\end{equation}
given by
\begin{equation}
     U_0(t,t_0)=
     \left[\frac{m}{2\pi i\hbar(t-t_0)}\right]e^{i\frac{m({\bf x}-{\bf x'})^2}{2\hbar(t-t_0)}}.
\end{equation}
We recognize here the so called interaction picture. On the basis of the duality principle
in perturbation theory as given in \cite{fra3}, one could ask what series is obtained when
the kinetic term is formally interchanged with the random potential in the Dyson series.
Indeed, this will give the series
\begin{eqnarray}
\label{eq:K}
      K(t,t_0) &=& K_0(t,t_0)\left[I+\frac{i}{\hbar}\int_{t_0}^t dt'
	    K_0^{-1}(t',t_0)\frac{\hbar^2}{2m}\Delta_2K_0(t',t_0)\right. \\ \nonumber
      &-& \frac{1}{\hbar^2}\left.\int_{t_0}^t dt' \int_{t_0}^{t'} dt''
      K_0^{-1}(t',t_0)\frac{\hbar^2}{2m}\Delta_2K_0(t',t_0)
      K_0^{-1}(t'',t_0)\frac{\hbar^2}{2m}\Delta_2K_0(t'',t_0)+\ldots\right]
\end{eqnarray}
being now $K_0$ the solution of the equation
\begin{equation}
     i\hbar\frac{\partial K_0}{\partial t}=V({\bf x})K_0
\end{equation}
being
\begin{equation}
\label{eq:K0}
    K_0(t,t_0) = e^{-\frac{i}{\hbar}V({\bf x})(t-t_0)}.
\end{equation}
This series is dual to the Dyson series\cite{fra3} and as shown in Ref.\cite{fra8} it
is a semiclassical series, the Wigner-Kirkwood expansion, that holds as a strong coupling
expansion. This series, as we showed in Ref.\cite{fra8}, has the same eigenvalue expansion
as the WKB series as it should be. We just showed that it is dual to the Dyson series
for this case.
  
Our aim will be to evaluate the leading and the first orders of this series in order to
have the analytical expression for the localization length and its first correction. Besides
the physical result, our aim is also to point out another approach to cope with problems
in the strong coupling limit that are quite common in condensed matter physics. So,
let us give explicitly the Wigner-Kirkwood expansion till first order, one has
\begin{eqnarray}
\label{eq:WK}
      K(t,0) &=& K_0(t,0)\left\{I-\left[\frac{it^3}{6m\hbar}(\nabla V)^2-
      t^2\left(\frac{1}{4m}\Delta_2V+
      \frac{i}{2m\hbar}\nabla V\cdot{\bf p}\right)+\frac{i}{\hbar}\frac{{\bf p}^2}{2m}t\right]+\ldots
      \right\}.
\end{eqnarray}

Firstly, let us consider the leading order. We take the average on the random potential to obtain
\begin{equation}
     \langle K_0(t,0)\rangle = \langle e^{-\frac{i}{\hbar}V({\bf x})t}\rangle
\end{equation}
that has the well-known result
\begin{equation}
     \langle K_0(t,0)\rangle = e^{-\frac{t^2}{\tau_\phi^2}}
\end{equation}
being
\begin{equation}
    \tau_\phi = \sqrt{4\pi\hbar\tau n(E)}
\end{equation}
giving a Gaussian decay in time with a time scale defined by $\tau_\phi$ that can be seen
as a coherence time. So, we can Fourier transform the unitary evolution operator to give
\begin{equation}
    \langle K_0(E)\rangle=\int_{-\infty}^{+\infty}e^{i\frac{E}{\hbar}t}\langle K_0(t,0)\rangle dt = 
	\sqrt{\pi}\tau_\phi e^{-\frac{E^2\tau_\phi^2}{4\hbar^2}}
\end{equation}
showing that the energy is strongly localized on a scale $\hbar/\tau_\phi$. So, we can get
immediately the localization length by the relation, in complete analogy with the 1D case \cite{mck},
\begin{equation}
    \frac{1}{\lambda(E)^2}=-\lim_{A\rightarrow\infty}\frac{1}{2A}\ln|\langle K_0(E)\rangle|^2
\end{equation}
giving finally
\begin{equation}
    \lambda(E) = \left(\frac{\hbar^3}{m\tau}\right)^{\frac{1}{2}}\frac{1}{E}.
\end{equation}
This is the most important result of the paper showing a scaling law of the
localization length with the energy similarly to numerical and theoretical
expectations \cite{ro1,ro2,rai} and with a critical exponent $\nu=1$. It is important
to note the confirmation of the singularity for $E=0$ while the critical exponent
in the scaling law may show differences that we are going to exploit computing
the next order correction as at the leading order we have an Anderson model with
pure disorder. We note anyhow that agreement with the Chayes criterion $\nu\ge 1$
is essential in our case of a very large disorder.
We emphasize that these results apply for off-diagonal disorder. In this case the singular behavior of the state at $E=0$ should be expected.

The next step is to compute the first order correction to this result through our Wigner-Kirkwood
expansion. In order to obtain this we use the fact that the random potential is defined
on a lattice and so the derivatives must be properly discretized. Then, we can compute the
averages. So, we consider the series (\ref{eq:WK}) and discretize taking for the derivative
\begin{equation}
    \nabla V = \left(\frac{V(x_k,y_k)-V(x_{k-1},y_k)}{a},\frac{V(x_k,y_k)-V(x_k,y_{k-1})}{a}\right).
\end{equation}
being $a$ the inter-particle spacing given by $a=1/\sqrt{\pi n}$ with $n$ the density of the two dimensional electron gas. By noting that the only terms surviving in the averages are those in the same site,
one gets the following
\begin{eqnarray}
    \langle V K_0(t,0)\rangle &=& e^{-\frac{t^2}{\tau_\phi^2}}\left(-i\frac{u_0^2}{\hbar}t\right) \\ \nonumber
    \langle (\nabla V)^2 K_0(t,0)\rangle &=& e^{-\frac{t^2}{\tau_\phi^2}}
    \left(\frac{2u_0^2}{a^2}-\frac{2u_0^4t^2}{a^2\hbar^2}\right) \\ \nonumber
    \langle \Delta_2V K_0(t,0)\rangle &=& e^{-\frac{t^2}{\tau_\phi^2}}\frac{4iu_0^2t}{a^2\hbar}.
\end{eqnarray}
The momentum operator can be managed by a proper choice of the initial electron field granting the substitution
of the operator itself with the eigenvalue for the successive computations. In any case we are considering the
kinetic term everywhere negligible even if its taking into account may displace the critical point from $E=0$
after Fourier transform to ${\bf p}^2/2m$. Finally, the averaged propagator can be written down as
\begin{equation}
    \langle K(t,0)\rangle = e^{-\frac{t^2}{\tau_\phi^2}}
    \left[1+u_0^2\left(\frac{p_x+p_y}{2ma\hbar^2}+\frac{2i}{3ma^2\hbar}\right)t^3
    +u_0^4\frac{i}{3ma^2\hbar^3}t^5+\ldots\right]
\end{equation}
that we can rewrite, introducing the Fermi time $\tau_F=ma^2/\hbar$, as
\begin{equation}
\label{eq:KA}
     \langle K(t,0)\rangle = e^{-\frac{t^2}{\tau_\phi^2}}
     \left[1+\frac{\tau_\phi}{\tau_F}\left(\frac{p_x+p_y}{\hbar}a+\frac{4}{3}i\right)
     \frac{t^3}{\tau_\phi^3}+\frac{4}{3}i\frac{\tau_\phi}{\tau_F}\frac{t^5}{\tau_\phi^5}+\ldots\right].
\end{equation}
We recognize two relevant aspects of this first order correction. One has a polynomial dependence on time
and then, at this stage, the scaling law we have obtained gets no corrections. Besides, we recognize a
development parameter given by $\tau_\phi/\tau_F$ that gives a working condition on the mean free time $\tau$
for our expansion to hold. It critically depends on the inter-particle spacing.

In order to verify our conclusion, we take the series (\ref{eq:KA}) and put the momentum to zero (pure disorder). Then, we take the Fourier transform obtaining
\begin{equation}
    \langle K(E)\rangle=\sqrt{\pi}\tau_\phi e^{-\frac{E^2\tau_\phi^2}{4\hbar^2}}
    \left[1+\frac{\tau_\phi}{\tau_F}P_1\left(\frac{E\tau_\phi}{\hbar}\right)+\ldots\right]
\end{equation}
where we have put
\begin{equation}
    P_1(x)=-\frac{7}{2}x+x^3-x^5. 
\end{equation}
The polynomial character of this correction changes to a logarithmic correction to the localization length that in
the limit $A\rightarrow\infty$ goes to zero preserving our initial result. It is interesting to point out that
what we have obtained is also a perturbation series in the development parameter $E_F/u_0$, being $E_F$ the
Fermi energy, and this is an asymptotic series in the strength of the disorder going to infinity as promised.

In conclusion we have applied the duality principle in perturbation theory to a two dimensional gas in presence of a strong disorder. This has shown an application of the semiclassical Wigner-Kirkwood expansion to a strongly perturbed system. We have obtained a series in the development parameter $E_F/u_0$ in agreement with expectations for a strongly disordered system. The localization length has been given explicitly with a power law scaling as $1/E$. This gives a critical exponent $\nu=1$ in agreement with the Chayes criterion $\nu\ge 1$. The method we applied is rather general and could be extensively applied to other problems in condensed matter physics.




\end{document}